\documentclass[aps,prl,reprint,amsmath,amssymb,longbibliography,nofootinbib]{revtex4-2}
\usepackage{amsmath,amssymb,graphicx,booktabs,bm}
\usepackage[T1]{fontenc}
\usepackage{lmodern}
\usepackage{xcolor}
\usepackage[breaklinks=true,colorlinks=true,linkcolor=blue!55!black,citecolor=blue!55!black,urlcolor=blue!55!black]{hyperref}
\usepackage{amsthm}

\newtheorem{corollary}{Corollary}
\newtheorem{proposition}{Proposition}

\begin{document}
\title{When does a noisy Bell pair attain the Rains bound?}
\author{Matthew Barber}
\author{Stefano Pirandola}
\affiliation{Department of Computer Science, University of York, York YO10 5GH, United Kingdom}
\begin{abstract}
Two entangled Bell-diagonal qubit states with the same singlet fraction have the same Rains bound. We prove that their attainable rates under operations preserving the positive partial transpose can nevertheless differ: the bound is attained if and only if one of the Bell probabilities vanishes. Every mixed entangled isotropic state, in every finite dimension, lies strictly below its bound. This refutes the attainment suggested by finite-block numerics, and places the two-way quantum capacity of the qubit depolarizing channel strictly below its relative-entropy bound.
\end{abstract}
\maketitle
Suppose a source supplies Bell pairs, each subject to a Pauli error that neither recipient can see. How much entanglement can they recover? For two qubits the most useful upper bound is remarkably economical: it depends only on the probability $F$ of the most likely Bell state, and so cannot tell a source with one kind of error from a source with three. With a single kind of error, a change of local basis reduces the problem to dephasing, and one-way hashing attains the bound \cite{bennett,rains}. We show that, if the parties may use operations preserving the positive partial transpose (PPT) \cite{rainsbound}, a second kind of error costs nothing further. A third, however rare, changes the answer.

The question concerns the asymptotic yield at vanishing error. PPT operations can do everything that local operations and classical communication (LOCC) can do, and for any finite block their best fidelity is given exactly by a single test \cite{rains}. Neither fact ensures that their asymptotic rate reaches the regularized Rains bound. Reference~\cite{lami} recently exhibited a two-qutrit Werner state for which it does not, using a converse based on analytic interpolation. Two-qubit Bell mixtures raise a different question: can one decide which members of a whole family attain the bound, even without knowing the rate itself?

\textit{Statement.---}Let $B_0,\ldots,B_3$ be the four Bell projectors, labelled so that $p_0=F$ is the largest probability, and let $\rho=\sum_i p_i B_i$. We write $D_{\rm PPT}$ for the largest number of ebits that can be distilled per input pair with fidelity tending to one, without preshared entanglement. The operations allowed are completely positive and trace preserving, and remain completely positive when conjugated by the partial transpose on Bob's systems. The state is separable for $F\le1/2$; on the entangled interval we prove the following classification:
\begin{align}
 D_{\rm PPT}(\rho)&=1-h_2(F) &&\text{if some }p_i=0,\nonumber\\
 D_{\rm PPT}(\rho)&<1-h_2(F) &&\text{if all }p_i>0.\label{eq:class}
\end{align}
Here $h_2$ is the binary entropy in bits. Consider, for example, the distributions $(3/4,1/8,1/8,0)$ and $(3/4,1/12,1/12,1/12)$. Their closest separable states give the same relative entropy, and from the first source PPT operations can distil all of it. From the second they cannot: there is an interval of rates just below that value at which even the best PPT protocol has exponentially small fidelity. A difference in performance on finite blocks would not establish this; the interval must survive as the block length tends to infinity.

The first case includes the pure and dephasing states. Exact two-qubit rates beyond the familiar maximally correlated form were obtained in Ref.~\cite{cornelio} (see also Ref.~\cite{coles}) under a separability condition on the complementary marginal, but that condition excludes rank-three two-qubit states. To our knowledge, the classification by support in Eq.~\eqref{eq:class}, including attainment on the whole rank-three face, is new. Since every two-qubit state with maximally mixed marginals is a Bell mixture up to local unitaries \cite{hh96}, Eq.~\eqref{eq:class} is also a rank criterion: such a state, if entangled, attains its Rains bound exactly when its rank is at most three. For unital qubit channels the corresponding condition is Kraus rank at most three \cite{kingruskai} (Appendix~\ref{app:rank}).

\begin{figure*}[t]
 \centering\includegraphics[width=.95\textwidth]{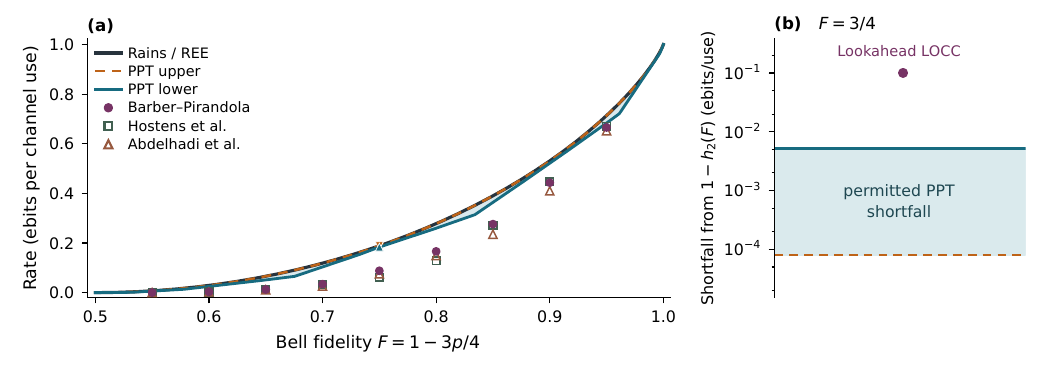}
 \caption[Entanglement rates for the depolarizing channel]{(a) Qubit depolarizing channel, for which the largest Bell probability of the Choi state is $F=1-3p/4$, with $p=1-\lambda$ the replacement parameter. On this scale the Rains/REE curve cannot be distinguished from the certified PPT upper bound. The lower PPT envelope combines certified block tests with hashing; the two triangles at $F=3/4$ mark the stronger fixed-state estimates. Circles show our numerical receding-horizon yields from Fig.~3 of Ref.~\cite{lookahead}, tabulated in Appendix~\ref{app:figure}; the other LOCC markers show the benchmarks of Refs.~\cite{hostens} ($q=6$) and~\cite{abdelhadi} from the same comparison. (b) Shortfall below the Rains bound at $F=3/4$. The PPT interval is certified; the LOCC marker is an achievable yield, not an upper bound.}\label{fig:depol}
\end{figure*}

\textit{The benchmark and its operational test.---}Put $e=1-F$ and $U=1-h_2(F)$ and, for $e>0$, write $\rho=F B_0+e\sigma$, with $\sigma=\sum_{i>0}v_i B_i$. The Rains bound is the infimum of the relative entropy $D(\rho\Vert\tau)$ over positive $\tau$ with $\|\tau^\Gamma\|_1\le1$, where $\Gamma$ denotes the partial transpose. This bound and the relative entropy to separable states both equal $U$, even after regularization \cite{vp,rains,audenaert}. From above, $(B_0+\sigma)/2$ is a mixture of the separable midpoints $(B_0+B_i)/2$, each describing complete dephasing in a Pauli basis, and it gives divergence $U$. From below, the positive score $X=2FB_0+2e(I-B_0)$ has $\|X^\Gamma\|_\infty=1$, so that $\operatorname{Tr}(X^{\otimes n}\tau)\le1$ for every Rains comparator on $n$ copies; measuring in the Bell basis and applying the classical relative-entropy inequality with this score gives the lower bound $n\operatorname{Tr}(\rho\log_2 X)=nU$. Comparator and score together settle the entropic optimization. An actual distillation protocol needs more.

The operational criterion of Ref.~\cite{rains} identifies the best fidelity with a maximally entangled output of dimension $K$ as the maximum of $\operatorname{Tr}(T\rho^{\otimes n})$ over
\begin{equation}
 0\le T\le I,\qquad \|T^\Gamma\|_\infty\le K^{-1}.       \label{eq:test}
\end{equation}
Behind this single operator lies a channel: given $T$, prepare the target Bell state $\Phi_K$ with probability $\operatorname{Tr}(T\rho^{\otimes n})$, and otherwise the normalized state on its orthogonal complement. Under partial transposition the output of this map splits into the symmetric and antisymmetric subspaces of the swap, and positivity on the two subspaces gives exactly the upper and lower bounds on $T^\Gamma$ in Eq.~\eqref{eq:test}. Conversely, averaging any distillation map over conjugate local unitaries on its output leaves its target fidelity unchanged and brings it to this form. Neither party needs to be told the outcome.

The same effect must therefore pass two quite different tests: its expectation on the input must be large, and its transposed norm must be small. The second requirement concerns the whole Hilbert space, including Bell labels on which the input has no weight.

\textit{Using an absent error.---}Suppose that $p_3=0$ and $1/2<F<1$, and put $s=e/F$. The three positive effects
\begin{equation}
 G=B_0+sB_3,\qquad A=B_1+B_2,\qquad Z=(1-s)B_3
\end{equation}
sum to the identity and have probabilities $F,e,0$. The term $sB_3$ changes the transposed norm of $G$ without changing these probabilities. The partial transposes of the Bell projectors are $B_i^\Gamma=I/2-\Pi_i$, where the $\Pi_i$ are orthogonal rank-one projections, and evaluating on these four projections gives the identity
\begin{equation}
                  2F|G^\Gamma|+2e|A^\Gamma|=I.            \label{eq:balance}
\end{equation}
Give the three outcomes the scores $2F,2e,0$, and on $n$ copies sum the product effects whose product of scores is at least $2^{nR}$. The sum is an effect, and its partial transpose lies between $\pm2^{-nR}I$: tensor \eqref{eq:balance} and apply the triangle inequality to each transposed eigenvalue. Its expectation on the input tends to one whenever $R<F\log_2(2F)+e\log_2(2e)=U$, so Eq.~\eqref{eq:test} yields a deterministic PPT protocol, with no rejected branch left uncharged. Only the final threshold test has to define a PPT map; the individual outcomes need not be readable by LOCC.

It helps to see where the freedom lies. On the support of the input, $G$ and $A$ merely distinguish an intact pair from one bearing an error. Many effects on the whole two-qubit space would give the same binary record, and one of them has exactly the transposed norm that the likelihood ratio requires. For this reason the construction is indifferent to how the total error probability is divided between $B_1$ and $B_2$. Delivering that binary record to the parties without disturbing the pairs would be a different, and generally stronger, requirement. Reference~\cite{wangduan} used freedom outside the input support in an earlier exact PPT distillation construction; here the same freedom suffices for every division of the error between the two remaining labels.

\textit{Why full support prevents attainment.---}A converse must hold for arbitrary collective effects, including those that keep no count of physical errors. Fix any $T$ satisfying \eqref{eq:test}. We work in natural logarithms, writing $\mathcal U=U\ln 2$, with $s=e/F$ as before, and introduce the polynomial
\begin{equation}
 P_n(z)=\exp(n\,\mathcal U)\operatorname{Tr}\!\left[T
               \left(\frac{B_0+z\sigma}{2}\right)^{\otimes n}\right].
                                                              \label{eq:poly}
\end{equation}
Its coefficient of $z^j$ is $2^{-n}\exp(n\,\mathcal U)\binom nj f_j$, where $0\le f_j\le1$ is the mean acceptance given $j$ errors; in particular the physical fidelity is $\mathbb E f_J$ with $J\sim{\rm Bin}(n,e)$. Polynomials of this kind for distillation tests go back to Ref.~\cite{rains}. We refine the analytic-family converse of Ref.~\cite{lami} (its Theorem~3): the pencil in \eqref{eq:poly} is not normalized, and the positivity of its coefficients supplies a second bound to set beside the complex partial-transpose estimate.

At $s$ the pencil is proportional to the physical state, and at $1$ it is the separable comparator. Hence $P_n(s)\le s^{ne}$ and $P_n(1)\le\exp(n\beta)$, where $\beta=\mathcal U-n^{-1}\ln K$. H\"older interpolation of the nonnegative coefficient sum, together with $|P_n(z)|\le P_n(|z|)$, shows that
\[
 u_n(z)=n^{-1}\ln|P_n(z)|-e\ln|z|
\]
is at most $\beta_+=\max(0,\beta)$ throughout $s\le|z|\le1$, and at most zero on the inner circle. The two real radii mean different things. The inner one gives a bound from ordinary positivity and the physical input alone, without appeal to the restriction on allowed operations; the outer one is a state that no admissible map can turn into a singlet with fidelity greater than $1/K$. On every intermediate circle the positive coefficients interpolate between the two. This interpolation by itself would only recover the usual entropic converse: strictness needs information from another direction in the complex plane. That information comes directly from \eqref{eq:test}, which gives a different upper bound at every complex $z\ne0$:
\begin{equation}
 u_n(z)\le\beta+\ln H(z)-e\ln|z|.            \label{eq:hnorm}
\end{equation}
Here $H(z)=\|[(B_0+z\sigma)/2]^\Gamma\|_1$. The tensor-product form of \eqref{eq:poly} is essential, for it makes the transposed trace norm exactly $H(z)^n$.

The difference between two errors and three shows itself at the point $z=-1$. Direct diagonalization gives
\begin{equation}
 H(z)=\tfrac14\left(|z-1|+\sum_{i=1}^3|1+(1-2v_i)z|\right).
\end{equation}
At $-1$ the four arguments are the real numbers $-2,2v_1,2v_2,2v_3$. If all $v_i>0$, then $H(-1)=1$ and the linear term about this point vanishes in every direction; indeed $H(-1+\xi)\le1+C|\xi|^2$ with $C=(\sum_i v_i^{-1}-7)/16$. If an error label is absent, one of the moduli has its corner exactly at this point, and the quadratic estimate fails.

To see how a tangency at a nonphysical point can cost a positive rate, average $u_n$ over a small circle of radius $b$ centred at a point of the unit circle near $-1$. On the inner part of this small circle use the annulus bound. On the outer part the last term of \eqref{eq:hnorm} falls to first order in $b$, whereas $\ln H$ rises at most to second order. The subharmonic mean-value inequality \cite{ransford} therefore gives $u_n\le\beta_+-\gamma$ on a whole arc of the outer circle, with $\gamma>0$ independent of the block length. For example, centres within angle $a$ of $-1$ and radius $0<b<1-s$ give
\[
 \gamma=eb/\pi-(C+e/8)b^2-Ca^2>0
\]
once $a=b$ is small enough. What matters is that the arc has nonzero length, since a gain at an isolated boundary point would leave the harmonic estimate unchanged. The quadratic inequality holds uniformly at every centre on the arc, because each lies within distance $a$ of $-1$. The constants are fixed once, from the source probabilities, before any block length or candidate rate is specified.

Seen from $z=\sqrt{s}$, this arc has positive harmonic measure: at least $\omega=a/[L\cosh(\pi^2/L)]$, where $L=\ln(1/s)$, while the whole outer circle has harmonic measure $1/2$. The two scales are clearest in the coordinate $z=s\exp(x+iy)$, in which the point of observation lies halfway across a strip of width $L$. There the density on the outer side is $[2L\cosh(\pi y/L)]^{-1}$, summed over angular periods, and the nearest period alone gives the stated lower bound. The two boundary estimates can therefore be averaged with weights that depend on the state but not on the test, and the maximum principle gives
\begin{equation}
             u_n(\sqrt{s})\le\beta_+/2-\gamma\omega.          \label{eq:dent}
\end{equation}
Conversely, the identity $P_n(r)/r^{ne}=\mathbb E[f_J(r/s)^{J-ne}]$, together with the concentration of $J$ about $ne$, forces $u_n(\sqrt{s})\ge-o(1)$ along any sequence of fidelities bounded away from zero. To see this, restrict to $|J-ne|\le n^{2/3}$: the discarded probability tends to zero, while on the retained counts the factor $(r/s)^{J-ne}$ costs at most a subexponential factor at fixed $r$. However complicated the effect $T$, this physical contribution rules out a negative constant on the right of Eq.~\eqref{eq:dent}. Hence $D_{\rm PPT}\le U-2\gamma\omega/\ln2$.

The gap may be small, but at every fixed input of full support it is uniform in $n$. Its dependence on the error probabilities matters: $C$ diverges as one of the $v_i$ tends to zero, and the useful arc may shrink with it. We therefore claim no fixed loss, independent of the source, from adding an arbitrarily weak third error, and indeed there is none: a one-parameter variant of the test behind Eq.~\eqref{eq:balance} gives $D_{\rm PPT}\ge U-O(\delta\log(1/\delta))$ for a third error of probability $\delta$ (Appendix~\ref{app:rank}). At any fixed rate above the upper bound the fidelity tends to zero exponentially. Appendix~\ref{app:converse} gives the norm and harmonic-measure estimates in full.

\textit{Isotropic states.---}In local dimension $d\ge2$, let $\Phi_d$ be the maximally entangled projector and $\sigma_d=(I-\Phi_d)/(d^2-1)$. The isotropic state $F\Phi_d+e\sigma_d$ is entangled precisely when $F>1/d$, and its additive Rains and separable relative-entropy bounds are
\[
 U_d=\log_2d-h_2(F)-e\log_2(d-1).
\]
The separable comparator is the isotropic state at $F=1/d$, the uniform unitary average of a conjugate product vector. The variable counted in a block is still binary, recording membership of the signal projector or its complement rather than the detailed error label; but the number of error labels changes both the comparator probabilities and the physical radius at which the pencil is evaluated. Replace the pencil above by $[\Phi_d+(d-1)z\sigma_d]/d$ and put $s=e/[F(d-1)]$. Since the swap has only two eigenspaces,
\[
 H_d(z)=\frac{|d+1+(d-1)z|+(d-1)|z-1|}{2d}.
\]
Again $H_d(-1)=1$, now with quadratic coefficient at most $(d-1)/8$, and the same argument proves $D_{\rm PPT}<U_d$ at \emph{every} $1/d<F<1$. In higher dimensions the isotropic states must be distinguished from the Werner states: the former are symmetric under conjugate local unitaries, the latter under equal ones. The qutrit separation of Ref.~\cite{lami} concerns the latter family.

Reference~\cite{fang} suggested, on the strength of finite-block calculations, that isotropic states attain the bound. That work used $d=3$, $F=0.9$, infidelity $10^{-3}$ and blocks of up to 100 copies. At that very state our certified asymptotic bounds are
\begin{equation}
    1.00344789\le D_{\rm PPT}<1.015863168\,.   \label{eq:fang}
\end{equation}
The upper endpoint lies below $U_3=1.0159669071\ldots$ ebits per copy. This does not contradict the reported linear programmes, which concern the fidelity at fixed error and finite block length; the conjectural step was the extrapolation from those to the asymptotic rate. Our proof of strictness in every dimension is analytic. Numerical search serves only to sharpen the bounds at particular states, and its results enter only after feasibility and rounding have been checked rigorously.

The two ends of the numerical interval mean different things. For the lower bound one prescribes a finite measurement on the error-count sectors of a block; its physical probabilities are nonnegative, but its transposed eigenvalues may have either sign. Repeating this measurement description over many blocks, and accepting only words with a sufficiently high total log score, gives an admissible global test. The physical probabilities fix its yield, and absolute bounds on the signed eigenvalues fix the threshold it can afford. At the two-qutrit point the witness uses 20 pairs and 12 outcomes; a larger block or a different set of effects might give a better rate. In the qubit calculation a further permutation of the input pairs allows different score bounds at different transposed row frequencies. The parties never need to observe these frequencies, and the number of copies used to specify the witness places no limit on the length of the distillation protocol.

The numerical upper bound keeps the same polynomial for an arbitrary code. In the logarithmic coordinate of the annulus, harmonic measure bounds the function on an intermediate vertical line by its bounds on the two neighbouring lines. Applying this inequality on a finite set of lines carries the deficit from the negative side of the complex plane towards the physical input. Every cell is enclosed in the conservative direction; refinement can improve the bound but is not needed for its validity. The finite witnesses and certificates are given in the ancillary files, and neither computation assumes that an optimizer has been found.

\textit{Channel consequences.---}The Pauli channel $\mathcal N(M)=\sum_i p_i P_i M P_i^\dagger$, $P_i\in\{I,X,Y,Z\}$, has Choi state $\rho$, up to the labelling of the Bell states. Teleportation through one copy of $\rho$ implements one use of the channel by LOCC, since each Pauli correction commutes through the noise up to a phase; conversely, one use on half of a locally prepared Bell pair distributes $\rho$. Substituting either procedure into a code, adaptive or not, gives $Q_{\rm PPT}(\mathcal N)=D_{\rm PPT}(\rho)$ and $Q_2(\mathcal N)=D_{\leftrightarrow}(\rho)$; for PPT assistance this is the reduction in Proposition~8 of Ref.~\cite{lm} (see also Ref.~\cite{plob}). A Pauli channel with one displacement missing therefore has PPT-assisted capacity exactly $1-h_2(F)$ if $F>1/2$, and zero otherwise. We do not assert equality under LOCC on a genuine rank-three face.

For the depolarizing channel the three error probabilities are all $e/3$; equivalently, $\mathcal N(M)=\lambda M+(1-\lambda)\operatorname{Tr}(M)I/2$ with $\lambda=(4F-1)/3$. Since every LOCC code is PPT preserving, the two-way quantum capacity obeys
\begin{equation}
 Q_2(\mathcal N)\le Q_{\rm PPT}(\mathcal N)<1-h_2(F)
            \qquad (1/2<F<1).                              \label{eq:chan}
\end{equation}
The right-hand side is the relative-entropy upper bound of Pirandola, Laurenza, Ottaviani and Banchi (PLOB) \cite{plob}. By contrast, PPT-assisted capacities attain the PLOB bound for thermal-loss, amplifier and additive-noise channels \cite{bosonic}. At the endpoints, the entanglement-breaking channel at $F=1/2$ and the noiseless channel at $F=1$ have capacities zero and one. At $F=3/4$ the computer-certified interval is $0.18355862\le Q_{\rm PPT}<0.188642407$, against $1-h_2(3/4)=0.1887218755\ldots$. The lower endpoint comes from a collective PPT test, not from a two-way LOCC protocol, and Fig.~\ref{fig:depol} keeps the two classes apart. At the same fidelity our lookahead protocol gives about $0.0880$ ebits per use (Ref.~\cite{lookahead}; tabulated in Appendix~\ref{app:figure}). At higher sampled fidelities the local protocols change order, so their yields must be compared at the same input noise. The separation proved in Eq.~\eqref{eq:chan} is from the converse curve, not from any of these lower bounds. Collective effects have improved rates before, in the 24-copy qutrit construction near Eq.~(131) of Ref.~\cite{lami}; the finite effects and certificates used here are for the isotropic problem.

\textit{Conclusion.---}The Rains bound of a Bell-diagonal qubit state depends on its largest probability alone; whether operations preserving the positive partial transpose can attain it depends on its support. When one Bell label is absent, a threshold test adjusted on the unoccupied subspace distils at exactly $1-h_2(F)$. When all four are present the bound is never attained, and above a rate strictly below it the fidelity of every protocol decays exponentially. The same strict inequality holds for every mixed entangled isotropic state in every finite dimension. This refutes the attainment suggested by finite-block numerics, and places the two-way quantum capacity of the depolarizing channel strictly below the PLOB bound throughout $1/2<F<1$. The converse rests on information that no single comparison state can supply: along a complex pencil the partial-transpose constraint meets the trivial bound at one nonphysical point only to second order, and harmonic measure carries the resulting deficit back to the physical input. The loss is small, lying between $7.9\times10^{-5}$ and $5.2\times10^{-3}$ ebits per pair at $F=3/4$, but it persists at every block length. Its closed form, the loss for unequal errors and the optimal two-way rates remain open.
\begin{acknowledgments}
This work was supported by UK Research and Innovation (UKRI) through the Integrated Quantum Networks Research Hub (IQN, EP/Z533208/1). OpenAI's ChatGPT (GPT-6 Astra model), a large language model, was used during the preparation of this manuscript to generate candidate mathematical derivations and proofs, to assist with their development and to improve the presentation of the manuscript. Anthropic's Claude (Opus 5.5) was also used to check derivations numerically, verify references and literature, and assist with preparing the manuscript. Both models were directed by prompts written by the authors concerning the mathematical assumptions, derivations and statements under consideration. All mathematical results, derivations and proofs produced with their assistance were independently examined, verified and revised as necessary by the authors, who take full responsibility for the accuracy and scientific content of the work.
\end{acknowledgments}

\clearpage
\onecolumngrid
\setcounter{secnumdepth}{2}
\appendix
\makeatletter\@addtoreset{equation}{section}\makeatother
\renewcommand{\theequation}{\thesection\arabic{equation}}
\renewcommand{\theHequation}{\thesection\arabic{equation}}
\setcounter{theorem}{2}\setcounter{corollary}{1}
\section{Summary of results}\label{app:summary}
\label{sm:summary}

For reference we collect here the statements proved in these appendices. Rates are measured in ebits per input pair, and $D_{\rm PPT}$ is the supremum of the rates at which completely PPT-preserving protocols reach fidelity tending to one. Throughout, $e=1-F$.

\textit{Theorem 1 (Bell-diagonal qubit states).} Let $\rho=\sum_{i=0}^3p_iB_i$ with $F=p_0=\max_ip_i>1/2$.
(a) If $p_i=0$ for some $i$, then $D_{\rm PPT}(\rho)=1-h_2(F)$.
(b) If $p_i>0$ for all $i$, put $v_i=p_i/e$, $C=(\sum_{i=1}^3v_i^{-1}-7)/16$, $s=e/F$ and $L=\ln(1/s)$. For every $0<b<1-s$ and $0<a<\pi$ with $\gamma=eb/\pi-(C+e/8)b^2-Ca^2>0$,
\[
 D_{\rm PPT}(\rho)\le1-h_2(F)-\frac{2a\gamma}{(\ln2)L\cosh(\pi^2/L)}<1-h_2(F),
\]
and at every rate above the middle expression the fidelity of any sequence of admissible tests decays exponentially in the block length. Such $a,b$ always exist.

\textit{Theorem 2 (isotropic states).} Let $d\ge2$, $1/d<F<1$ and $\rho=F\Phi_d+e(I-\Phi_d)/(d^2-1)$. The same bound and strong converse hold with $1-h_2(F)$ replaced by $U_d=\log_2d-h_2(F)-e\log_2(d-1)$, with $C=(d-1)/8$ and $s=e/[F(d-1)]$.

\textit{Corollary 1 (qubit Pauli channels).} If $\mathcal N$ has Choi state $\rho$ as in Theorem~1, then $Q_{\rm PPT}(\mathcal N)=D_{\rm PPT}(\rho)$. In particular $Q_2(\mathcal N)\le Q_{\rm PPT}(\mathcal N)<1-h_2(F)$ for the depolarizing channel at every $1/2<F<1$.

Theorem~1(a) is proved in Appendix~\ref{app:absent}; Theorem~1(b) and Theorem~2 in Appendix~\ref{app:converse}, culminating in Eq.~\eqref{sm:strict}; and Corollary~1 in Appendix~\ref{app:pauli}. In both theorems the quantity $1-h_2(F)$ or $U_d$ on the right equals the Rains bound, the relative entropy of entanglement, and their regularizations, as we show next.

\textit{Further results.} Appendix~\ref{app:rank} restates Theorem~1 as a condition on rank, for two-qubit states with maximally mixed marginals and for unital qubit channels (Corollary~\ref{cor:rank}), and shows that the rate is nonetheless continuous at the rank-three face (Proposition~\ref{prop:face}). Appendices~\ref{app:cert} and~\ref{app:figure} give the computer-certified sharpenings at two particular states and explain how the figure was constructed.

\section{Operational test and additive comparison bound}\label{app:test}
\label{sm:oper}

The partial transpose $\Gamma$ is taken on every system held by Bob. A bipartite channel $\Lambda$ is completely PPT preserving if $\Gamma_{\rm out}\Lambda\Gamma_{\rm in}$ is completely positive. We impose no non-signalling condition and provide no shared entanglement. Fidelity means overlap with the target projector, and rates are measured in ebits per input pair. Set
\begin{equation}
 |\Omega_d\rangle=d^{-1/2}\sum_{j=0}^{d-1}|j,j\rangle,\qquad
 \Phi_d=|\Omega_d\rangle\langle\Omega_d|.
\end{equation}
The fidelity theorem of Ref.~\cite{rains}, Theorem~3.1, states that the greatest attainable overlap with $\Phi_K$ is
\begin{equation}
 \max_T\left\{\operatorname{Tr}(T\rho^{\otimes n}):
 0\le T\le I,\quad -I/K\le T^\Gamma\le I/K\right\}.
 \label{sm:rains}
\end{equation}
To see this, note first that averaging the output of a code under $U\otimes\bar U$ preserves its overlap with $\Phi_K$. The averaged channel prepares either $\Phi_K$ or $(I-\Phi_K)/(K^2-1)$, with effects $T$ and $I-T$. After conjugation by the partial transpose its Choi matrix splits into two output swap sectors, whose coefficients are, up to the input transpose, $(I+KT^\Gamma)/[K(K+1)]$ and $(I-KT^\Gamma)/[K(K-1)]$. Their positivity is precisely the condition in~\eqref{sm:rains}, and conversely every such effect defines a channel of this form. Here $K\ge2$, since a target of size one is immaterial. The effect in~\eqref{sm:rains} therefore certifies a trace-preserving protocol, whose complement is part of the protocol rather than a discarded outcome.

Let $e=1-F$, and let $\Phi$ stand for $\Phi_d$. In these appendices we consider one of two cases:
(i) $d=2$ and $\rho=F\Phi+e\sigma$ is Bell diagonal, with $\Phi$ its most probable label and $\sigma$ supported on the other three labels; or (ii) $d\ge2$ and $\sigma=(I-\Phi)/(d^2-1)$, with $F$ the isotropic singlet fraction. We first suppose $1/d<F<1$, and write
\begin{equation}
 U_d=\log_2d-h_2(F)-e\log_2(d-1),\quad
 \tau=\frac{\Phi+(d-1)\sigma}{d}.
 \label{sm:benchmark}
\end{equation}
In both cases $\tau$ is separable. For qubits, writing $\sigma=\sum_{i=1}^{3}v_i B_i$ gives $\tau=\sum_iv_i(\Phi+B_i)/2$, and each summand is the Choi state of a complete dephasing, an equal mixture of two conjugate product vectors in the appropriate Pauli basis. For isotropic states, the Haar twirl of $|\bar u,u\rangle\langle\bar u,u|$ is isotropic with singlet fraction $1/d$, and so equals $\tau$. Since $\sigma\Phi=0$, a direct calculation gives $D(\rho\Vert\tau)=U_d$, with $D$ in binary logarithms.

This trial state gives an upper bound; we now show that $U_d$ is also a lower bound on every regularized comparison. The Rains relative entropy minimizes $D(\rho\Vert\omega)$ over $\omega\ge0$ with $\|\omega^\Gamma\|_1\le1$, allowing subnormalization. Put
\begin{equation}
 G_*=dF\Phi+\frac{de}{d-1}(I-\Phi).
\end{equation}
Since $d\Phi^\Gamma$ is the swap, the eigenvalues of $G_*^\Gamma$ are $1$ and $(d+1)e/(d-1)-F\in[-1,1]$, so every Rains comparison operator on $n$ pairs obeys $\operatorname{Tr}(G_*^{\otimes n}\omega)\le1$. Measure the two sectors $\Phi,I-\Phi$ on each pair, and let $g$ be the product of the corresponding eigenvalues of $G_*$. The classical log-sum inequality gives $D(P\Vert Q)\ge\sum P\log_2g-\log_2\sum Qg$, and data processing then bounds $D(\rho^{\otimes n}\Vert\omega)$ below by $nU_d$. The reverse inequality follows from $\tau^{\otimes n}$. Hence the Rains and separable relative entropies, and their regularizations, all equal~\eqref{sm:benchmark}, and the usual Rains converse bounds the PPT rate by this value. The same vanishing-error converse also follows directly: for any feasible $T$, $\operatorname{Tr}(T\tau^{\otimes n})\le1/K$, and binary data processing does the rest.

\section{An absent Bell label}\label{app:absent}
\label{sm:face}

In case (i), suppose that one of the non-leading labels is absent. After relabelling, write $B_0=\Phi$, $p_3=0$ and $s=e/F$, and consider the three positive effects
\begin{equation}
 G=B_0+s B_3,\qquad A=B_1+B_2,\qquad Z=(1-s)B_3 .
\end{equation}
They sum to $I$, and their probabilities on the physical state are $F,e,0$. Assign the scores $2F,2e,0$ to $G,A,Z$ respectively. We claim the exact operator identity
\begin{equation}
       2F|G^\Gamma|+2e|A^\Gamma|=I.\label{sm:score}
\end{equation}
The partial transposes share a common eigenbasis: $B_i^\Gamma=I/2-P_i$, where the four rank-one projections $P_i$ form an orthogonal resolution of the identity. This follows from $2B_0^\Gamma=V$, the swap, whose negative eigenspace is spanned by the singlet; the four local Pauli translates of the singlet are orthogonal. On the $P_0,P_3$ rows $|A^\Gamma|=1$ and $|G^\Gamma|=(1-s)/2$, while on the $P_1,P_2$ rows the corresponding values are $0$ and $(1+s)/2$. This proves~\eqref{sm:score}.

Let $T_n$ be the sum of those tensor products of these effects whose product score is at least $2^{nR}$. As a sum of outcomes of a product positive operator-valued measure, it satisfies $0\le T_n\le I$. For each retained outcome string, the absolute value of the transposed product is the tensor product of the corresponding absolute values. Bounding each transposed term above and below by its absolute value and summing gives
\begin{equation}
 \pm T_n^\Gamma\ \le\ 2^{-nR}
    \bigl(2F|G^\Gamma|+2e|A^\Gamma|\bigr)^{\otimes n}
    = 2^{-nR}I .
\end{equation}
Equation~\eqref{sm:rains} therefore applies with $K=\lfloor2^{nR}\rfloor$, and by the law of large numbers the acceptance tends to one whenever $R<F\log_2(2F)+e\log_2(2e)=U_2$. The three auxiliary outcomes need not be measurable by PPT operations on single copies; only the final threshold effect has to satisfy~\eqref{sm:rains}. Altering an effect outside the support has a precedent in Proposition~5 of Ref.~\cite{wangduan}, for a different, two-qutrit construction. Here the score identity gives the asymptotic rate on every qubit Bell face.

Exact rates for mixed two-qubit states were known before this work, and need not come from an ordinary maximally correlated subspace. Theorem~2 of Ref.~\cite{cornelio} obtains the coherent-information rate when the complementary marginal $AC$ of a purification $ABC$ is separable, with the regularized separable relative entropy equal to that rate; Equation~(41) of Ref.~\cite{coles} restates this equality. The LOCC and relative-entropy bounds then trap the PPT rate as well. That condition cannot hold on the present rank-three face, for $\operatorname{rank}(AC)=\operatorname{rank}(B)\le2$ whereas $\operatorname{rank}(C)=\operatorname{rank}(AB)=3$, and the marginal of a separable state never has larger rank than the state itself. To see the last point, choose from a separable decomposition independent product vectors spanning the range, and take the span of their local factors.

The endpoints are straightforward. At $F=1$ the input is itself maximally entangled. At $F\le1/2$ a two-qubit Bell-diagonal state is separable, because the polytope of probabilities with $p_i\le1/2$ has as its vertices the equal mixtures of two Bell labels, all of which are separable. An isotropic state is separable for $F\le1/d$, as one sees by interpolating between the isotropic twirls of an orthogonal product and of $|\bar u,u\rangle$. These endpoints are set aside from the mixed entangled cases treated below.

\section{An analytic converse uniform in the block length}\label{app:converse}
\label{sm:annulus}

We prove the assertions for full-support qubit states and for mixed entangled isotropic states together. The proof refines the analytic-family method of Ref.~\cite{lami}, Theorem~3. That theorem uses a trace-one analytic family and integrates the lesser of two boundary trace-norm estimates. Here the pencil need not be normalized away from the positive comparison points. Positivity of the coefficients gives a bound throughout the annulus, and averaging this bound together with a transposed-norm estimate outside the annulus yields a strict deficit on part of its boundary. These are additional steps rather than a direct substitution in Theorem~3. The polynomial representation itself goes back to the test polynomials of Ref.~\cite{rains}, Theorem~5.4.

Write $\mathcal U=U_d\ln2$ in nats, and set
\begin{align}
 s&=\frac{e}{F(d-1)}\in(0,1),&L&=\ln(1/s),\nonumber\\
 Z(z)&=\frac{\Phi+(d-1)z\sigma}{d},&
 P_n(z)&=e^{n\mathcal U}\operatorname{Tr}\bigl[T Z(z)^{\otimes n}\bigr].
 \label{sm:pencil}
\end{align}
The symbol $e$ without an exponent denotes the error probability, and $e^{x}=\exp x$. Expanding the tensor power and averaging the acceptance of $T$ over the positions of the error factors gives, for each weight $j$, a number $f_j\in[0,1]$, and
\begin{equation}
 P_n(z)=\frac{\exp(n\mathcal U)}{d^n}
 \sum_{j=0}^n {n\choose j}(d-1)^j f_j z^j .
 \label{sm:polynomial}
\end{equation}
No symmetrization of the code is needed. If $J$ is binomial with parameters $(n,e)$, the physical fidelity is $\mathbb E f_J$; more generally, for every real $r>0$,
\begin{equation}
 r^{-ne}P_n(r)=\mathbb E[f_J(r/s)^{J-ne}].
 \label{sm:tilt}
\end{equation}
This follows on inserting the binomial probabilities and using $\exp(\mathcal U)/(dF)=s^e$.

Fix any feasible $(K,T)$ in~\eqref{sm:rains}, and write $\beta=\mathcal U-(\ln K)/n$ and $\beta_+=\max(0,\beta)$. Leaving aside the zero polynomial, which already has zero fidelity, define
\begin{equation}
 u_n(z)=n^{-1}\ln|P_n(z)|-e\ln|z|,\quad z\ne0.
\end{equation}
It is subharmonic~\cite{ransford}, taking the value $-\infty$ at zeros. Equation~\eqref{sm:tilt} gives $P_n(s)\le s^{ne}$. At $z=1$, $Z(1)=\tau$ is PPT, so $P_n(1)\le\exp(n\beta)$. For $r=s^t$ with $0\le t\le1$, positivity of all the coefficients and H\"older's inequality give $|P_n(z)|\le P_n(r)\le P_n(s)^tP_n(1)^{1-t}$ on $|z|=r$. Thus
\begin{equation}
 u_n\le\beta_+\ \ (s\le|z|\le1),\qquad
 u_n\le0\ \ (|z|=s).
 \label{sm:radial}
\end{equation}
A second inequality, valid for every $z\ne0$, follows because the bilinear trace pairing is invariant under partial transpose:
\begin{equation}
 u_n(z)\le\beta+\ln h(z)-e\ln|z|,\qquad
 h(z)=\|Z(z)^\Gamma\|_1.\label{sm:complex}
\end{equation}
Here we use that the operator norm of $T^\Gamma$ is at most $1/K$, and that the trace norm is multiplicative on tensor products even when $Z(z)^\Gamma$ is not Hermitian.

\subsection{The norm at the negative point}

For qubits we use the common eigenspaces $P_i$ described above. If $\sigma=\sum_{i=1}^3v_i B_i$, one obtains
\begin{equation}
 h(z)=\frac{|z-1|+\sum_{i=1}^3|1+(1-2v_i)z|}{4}.
 \label{sm:hnorm}
\end{equation}
If all $v_i>0$, the arguments of these moduli at $z=-1$ are $-2,2v_1,2v_2,2v_3$. For real $b\ne0$ and complex $t$, concavity of the square root gives
\begin{equation}
 |b+t|\le |b|+\operatorname{sgn}(b)\operatorname{Re}t
                          +|t|^2/(2|b|).
\end{equation}
The linear contributions cancel because $\sum_i(1-2v_i)=1$, and it follows that
\begin{equation}
 h(-1)=1,\qquad
 h(-1+\xi)\le1+C|\xi|^2,\qquad
 C=\tfrac1{16}\biggl(\sum_i v_i^{-1}-7\biggr).
 \label{sm:quadratic}
\end{equation}
A missing label produces a corner at this point and invalidates the quadratic argument.

For the $d$-dimensional isotropic state, $Z(z)^\Gamma$ is scalar on each of the two swap eigenspaces, with eigenvalues $[d+1+(d-1)z]/[d^2(d+1)]$ and $(z-1)/d^2$ and multiplicities $d(d+1)/2$ and $d(d-1)/2$ respectively. Therefore
\begin{equation}
 h(z)=\frac{|d+1+(d-1)z|+(d-1)|z-1|}{2d}.
 \label{sm:isoh}
\end{equation}
At $z=-1$ the real arguments are $2,-2$, and the same cancellation gives~\eqref{sm:quadratic} with $C=(d-1)/8$. For $d=2$ and $v_i=1/3$ the two calculations agree.

\subsection{A deficit of positive harmonic measure}

Choose $0<b<1-s$ and $0<a<\pi$ such that
\begin{equation}
 \gamma=\frac{eb}{\pi}-(C+e/8)b^2-Ca^2>0.
 \label{sm:gamma}
\end{equation}
This is always possible: take $a=b$ and then make $b$ small. Using~\eqref{sm:radial} inside the outer circle and~\eqref{sm:complex}--\eqref{sm:quadratic} outside it, we obtain, for $|z|\ge s$,
\begin{equation}
 u_n(z)\le\beta_+ +C|1+z|^2-e\ln^+|z|,
 \qquad \ln^+r=\max(0,\ln r).
 \label{sm:glue}
\end{equation}
Take a point $c=-\exp(i\theta)$ with $|\theta|\le a$, and average over the circle $z=c(1+b\exp(it))$, whose closed disc lies outside the inner circle. The mean of $|1+z|^2$ is $|1+c|^2+b^2\le a^2+b^2$. On the half-circle $\cos t\ge0$ we have $|1+b\exp(it)|\ge1+b\cos t$, and elsewhere $\ln^+|1+b\exp(it)|\ge0$; integrating $\ln(1+x)\ge x-x^2/2$ over the first half then gives a mean of at least $b/\pi-b^2/8$. Applying the subharmonic mean-value inequality to~\eqref{sm:glue} proves
\begin{equation}
        u_n(-\exp(i\theta))\le\beta_+-\gamma,\qquad |\theta|\le a .
 \label{sm:dent}
\end{equation}
In particular, the saving does not tend to zero with $n$.

Lift $s<|z|<1$ to the strip $z=s\exp(x+iy)$, $0<x<L$, with $y$ periodic of period $2\pi$. Seen from $(L/2,0)$, the harmonic measure on the side $x=L$ has density
\begin{equation}
 p_L(y)=\frac1{2L\cosh(\pi y/L)} .
\end{equation}
To check the normalization, note that $\exp[i\pi(x+iy)/L]$ maps the strip onto the upper half-plane and the observation point to $i$. On the negative real axis the Cauchy density $dt/[\pi(1+t^2)]$ becomes $p_L(y)\,dy$, so this side has mass $1/2$. Summing over $2\pi$-translates gives the kernel of the annulus, whose mass on the arc in~\eqref{sm:dent} is at least
\begin{equation}
             \omega=\frac{a}{L\cosh(\pi^2/L)} .
\end{equation}
Indeed, the part of the arc lying in $[-\pi,\pi]$ has total angular length $2a$, and there the unperiodized density is at least $p_L(\pi)$. The subharmonic maximum principle, with~\eqref{sm:radial} on the inner circle and~\eqref{sm:dent} on the outer arc, then yields
\begin{equation}
                u_n(\sqrt{s})\le\beta_+/2-\omega\gamma .
 \label{sm:propagate}
\end{equation}
To use continuous boundary data one may shrink the arc slightly before approximating its indicator. Zeros of $P_n$ cause no difficulty in an upper comparison.

\subsection{Exponential rejection at the source}\label{app:rejection}

In bits, the conclusion is the uniform strong-converse bound
\begin{equation}
 D_{\mathrm{PPT}}(\rho)\le
 U_d-\frac{2a\gamma}{(\ln 2)L\cosh(\pi^2/L)}<U_d.
 \label{sm:strict}
\end{equation}
Indeed, if $\liminf_n n^{-1}\log_2K_n$ lies strictly above the right-hand side, then~\eqref{sm:propagate} gives $u_n(r)\le-c_0$ at $r=\sqrt s$, for some fixed $c_0>0$ and all large $n$. Since $r/s>1$,~\eqref{sm:tilt} implies, for $0<t<\min(e,c_0/\ln(r/s))$,
\begin{equation}
 \mathbb Ef_J\le
 \Pr\{J<n(e-t)\}
 +\exp[-n(c_0-t\ln(r/s))].
\end{equation}
By a binomial Chernoff bound the first term also decays exponentially. This excludes every sequence of admissible tests at a rate above~\eqref{sm:strict}, not merely a proposed family of optimizers.

Reference~\cite{fang} solved a finite-block optimization for the isotropic state with $d=3$, $F=9/10$, at infidelity $10^{-3}$; its inference about the limit was an extrapolation, offered without proof. Here $L=\ln18$ and $C=1/4$, and the choice $b=3/50$, $a=1/28$, for example, gives $\gamma>0$ in~\eqref{sm:gamma}. The very state used in that test therefore refutes the extrapolation by an analytic inequality, without any search over effects. The asymptotic numerical enclosures given below are independent of the fixed-error data of Ref.~\cite{fang}.

\section{Transfer to Pauli channels}\label{app:pauli}

With the normalized Choi convention, $\sum_i p_i B_i$ is the Choi state of $\mathcal N(M)=\sum_i p_i W_i M W_i^\dagger$ for the four qubit Pauli operators. Since the Pauli operators commute up to phases, each Pauli correction passes through $\mathcal N$, and teleportation over the Choi state simulates one use exactly by LOCC. Conversely, one use on half of a locally prepared Bell pair supplies the Choi state. Substituting into an arbitrary adaptive protocol gives
$Q_{\rm PPT}(\mathcal N)=D_{\rm PPT}(J_{\mathcal N})$ and
$Q_2(\mathcal N)=D_{\leftrightarrow}(J_{\mathcal N})$; for PPT assistance this is the reduction of Ref.~\cite{lm}, Proposition~8. The Rains value is therefore the PPT capacity whenever an entangled qubit Pauli channel has a missing probability, and~\eqref{sm:strict} applies whenever all the probabilities are present. In particular $Q_2\le Q_{\rm PPT}<U_2$ for every mixed entangled qubit depolarizing Choi state, where $U_2$ is precisely the relative-entropy bound of Ref.~\cite{plob}. The pure and entanglement-breaking endpoints were treated at the end of Appendix~\ref{app:absent}.

\section{Rank criterion and continuity at the rank-three face}\label{app:rank}
\label{sm:rank}

\subsection{Statement in terms of rank}

\begin{corollary}[rank criterion]\label{cor:rank}
Let $\rho$ be an entangled two-qubit state with maximally mixed marginals and largest eigenvalue $\lambda$. Then its Rains bound is $1-h_2(\lambda)$, and $D_{\rm PPT}(\rho)=1-h_2(\lambda)$ if and only if $\operatorname{rank}\rho\le3$. Let $\mathcal N$ be a unital qubit channel whose Choi state has largest eigenvalue $\lambda>1/2$. Then $Q_{\rm PPT}(\mathcal N)=1-h_2(\lambda)$ if $\mathcal N$ has Kraus rank at most three, and $Q_2(\mathcal N)\le Q_{\rm PPT}(\mathcal N)<1-h_2(\lambda)$ if it has Kraus rank four.
\end{corollary}

\begin{proof}
A two-qubit state with maximally mixed marginals is $\tfrac14(I+\sum_{jk}t_{jk}\sigma_j\otimes\sigma_k)$. Local unitaries act on the real matrix $(t_{jk})$ by proper rotations on either side, so they bring it to diagonal form, and the state to a mixture of the four Bell projectors~\cite{hh96}. Local unitaries change neither the class of completely PPT-preserving operations, nor the Rains bound, nor the spectrum. A Bell mixture is entangled exactly when its largest probability exceeds $1/2$, and has a vanishing probability exactly when its rank is at most three. The first assertion is therefore a restatement of Theorem~1. A unital qubit channel is a Pauli channel preceded and followed by unitaries~\cite{kingruskai}; these change neither $Q_{\rm PPT}$ nor $Q_2$, and the Kraus rank is the rank of the Choi state. The second assertion then follows from Corollary~1 together with Theorem~1.
\end{proof}

\subsection{The rate is continuous at the face}

Attainment fails as soon as the fourth Bell probability is positive, but the rate itself does not jump.

\begin{proposition}[lower bound near the face]\label{prop:face}
Let $\rho=\sum_ip_iB_i$ with $F=p_0>1/2$, $e'=p_1+p_2$ and $\delta=p_3>0$. For every $0<t<1$, with $q=(1-t)\delta$,
\begin{equation}
 D_{\rm PPT}(\rho)\ \ge\ (F+t\delta)\log_2\frac{2}{1+t}+e'\log_2\frac{2t}{1+t}
 +q\log_2(2\delta)+(1-q)\log_2(1-q).
 \label{sm:facebound}
\end{equation}
\end{proposition}

\begin{proof}
Take the effects of Appendix~\ref{app:absent} with a free parameter, $G=B_0+tB_3$, $A=B_1+B_2$, $Z=(1-t)B_3$, but now give $Z$ a positive score. In the common eigenbasis of the partial transposes, on the rows $P_0,P_3$ one has $|G^\Gamma|=|Z^\Gamma|=(1-t)/2$ and $|A^\Gamma|=1$; on the rows $P_1,P_2$ one has $|G^\Gamma|=(1+t)/2$, $|A^\Gamma|=0$ and $|Z^\Gamma|=(1-t)/2$. Take the scores
\begin{equation}
 w_G=\frac{2(1-q)}{1+t},\qquad w_A=t\,w_G,\qquad w_Z=2\delta .
\end{equation}
Both row sums equal $(1-q)+q=1$, so $w_G|G^\Gamma|+w_A|A^\Gamma|+w_Z|Z^\Gamma|=I$. The outcome probabilities on $\rho$ are $F+t\delta$, $e'$ and $q$. The argument following Eq.~\eqref{sm:score} applies without change: the sum $T_n$ of the product effects whose product of scores is at least $2^{nR}$ satisfies $\pm T_n^\Gamma\le2^{-nR}I$, and its acceptance tends to one for every $R$ below the mean logarithmic score. That mean is the right-hand side of~\eqref{sm:facebound}.
\end{proof}

Fix $F$ and a ratio $p_1/p_2\in(0,\infty)$, so that the limiting face has rank three, and let $\delta\to0$. Taking $t=e'/F$ in~\eqref{sm:facebound} gives $D_{\rm PPT}\ge U_2-(1-t)\,\delta\log_2(1/\delta)-O(\delta)$. In Theorem~1(b) the constant $C$ is of order $e/\delta$, so admissible $a,b$ are of order $\delta$ and the bound reads $D_{\rm PPT}\le U_2-c\,\delta^2$ for a constant $c>0$ depending on $F$. Hence
\begin{equation}
 U_2-O\bigl(\delta\log(1/\delta)\bigr)\ \le\ D_{\rm PPT}(\rho)\ \le\ U_2-\Omega(\delta^2),
\end{equation}
and $D_{\rm PPT}\to U_2$ as the third error disappears. Table~\ref{tab:face} illustrates this at $F=3/4$ with $p_1=p_2$. Even at the isotropic point the single-copy test gives $0.069$ ebits, where hashing gives nothing. The true order of the shortfall, somewhere between $\delta^2$ and $\delta\log(1/\delta)$, is not known.

\begin{table}[h]
\caption{Floating-point evaluations of the analytic bounds near the rank-three face at $F=3/4$, $p_1=p_2$, where $U_2\simeq0.188722$. The second column numerically optimizes Eq.~\eqref{sm:facebound} over $t$; the last numerically optimizes the shortfall bound of Theorem~1(b) over $a$ and $b$. Values are rounded to the nearest digit shown; they illustrate the proved inequalities but are not one-sided certificates.}\label{tab:face}
\begin{ruledtabular}
\begin{tabular}{lccc}
$\delta=p_3$ & lower bound~\eqref{sm:facebound} & $U_2-{}$lower bound & shortfall bound (approx.)\\
\hline
$1/12$ (isotropic) & 0.068990 & $1.2\times10^{-1}$ & $7.3\times10^{-7}$\\
0.05 & 0.092578 & $9.6\times10^{-2}$ & $3.6\times10^{-7}$\\
0.02 & 0.132492 & $5.6\times10^{-2}$ & $3.9\times10^{-8}$\\
0.01 & 0.153829 & $3.5\times10^{-2}$ & $8.0\times10^{-9}$\\
0.001 & 0.182990 & $5.7\times10^{-3}$ & $6.7\times10^{-11}$\\
\end{tabular}
\end{ruledtabular}
\end{table}

\section{Numerical certificates for the two isotropic examples}\label{app:cert}
\label{sec:certificates}
Here $d$ is the local dimension, $F$ the singlet fraction and $e=1-F$,
and rates refer to vanishing error per original input pair. Both
numerical conclusions rest on feasible witnesses and one-sided
enclosures, with no extrapolation in the block length; the computer
certifies only their decimal evaluation. The data specifying the
witnesses, and independent checks of them, are provided as
ancillary files.

\subsection{A lower bound from finite blocks}
On $m$ copies, let $i$ count the occurrences of $\Phi_d$, and prescribe
a finite auxiliary measurement by numbers $a_y(i)\ge0$ with
$\sum_y a_y(i)=1$ on every count sector. Its probabilities and signed
partial-transpose eigenvalues are
\begin{align}
 p_y&=\sum_{i=0}^m {m\choose i}F^i e^{m-i}a_y(i),\\
 Q_j(y)&=d^{-m}\sum_{i=0}^m a_y(i)[t^i](d-1+t)^{m-j}(d+1-t)^j .
 \label{eq:cert-signed}
\end{align}
These follow from $\Phi_d^\Gamma=V/d$, where $V$ is the swap and $j$
counts the antisymmetric swap sectors. The $Q_j(y)$ may be negative,
but for each $j$ they sum over $y$ to one.

Consider a finite partition of $[0,1]$ into closed intervals $I$ with
disjoint interiors, and for positive scores $v_I(y)$ put
\begin{align}
 L_I(j)&=\sum_y v_I(y)|Q_j(y)|,\notag\\
 A_I(t)&=\sum_j{m\choose j}t^j(1-t)^{m-j}L_I(j).
\end{align}
Then the following rate, in bits, is achievable:
\begin{equation}
 {1\over m}\min_I\left\{\sum_y p_y\log_2 v_I(y)
             -\log_2\sup_{t\in I}A_I(t)\right\}.\label{eq:cert-type}
\end{equation}
To prove this, take $B$ independent blocks, and retain each outcome
word that passes \emph{all} of the score cutoffs
$\sum_{b=1}^B\log_2 v_I(y_b)\ge B(\sum_y p_y\log_2v_I(y)-\eta)$,
where $\eta>0$. By the law of large numbers the resulting effect has
acceptance tending to one. Average it over permutations of all $mB$
individual pairs, including permutations that move pairs between
blocks. In a transposed sector with $J$ antisymmetric sites, choose an
interval $I$ containing $t=J/(mB)$. Taking absolute values and dropping
all the other cutoffs bounds the eigenvalue in modulus by
\begin{equation}
 2^{-B(\sum_y p_y\log_2v_I(y)-\eta)}
 \;\mathbb E\left[\prod_b L_I(J_b)\ \Big|\ \sum_bJ_b=J\right],
\end{equation}
where, before conditioning, the $J_b$ are independent binomials with
parameters $(m,t)$. The conditional distribution is exactly that of the
permuted sector, not an approximation to it. The conditioning event is a
mode of $\mathrm{Bin}(mB,t)$ and so has probability at least
$(mB+1)^{-1}$, also at the two endpoints; the conditional expectation
is therefore at most $(mB+1) A_I(t)^B$. The test criterion of
Ref.~\cite{rains} now gives Eq.~\eqref{eq:cert-type}, on letting $B$
tend to infinity and then $\eta$ to zero. The auxiliary measurement is
simply a description of a single admissible global effect: its outcomes
are not assumed to be locally available, or free to discard. Without the
permutation or the partition, the same threshold argument gives the
slightly simpler lower bound
$[\sum_y p_y\log_2v(y)-\log_2\max_j L(j)]/m$. The weighted-effect
bounds and collective Werner-state examples of Ref.~\cite{lami} precede
this construction.

At $(d,F)=(2,3/4)$ the witness has $m=94$, 51 outcomes and ten dyadic
intervals. The $a_y(i)$ have common denominator $2^{192}$, and on each
interval $v_I(y)=2^{s_I(y)/64}$ with integer $s_I(y)$. The rational
arrays satisfy positivity and normalization exactly. To bound $A_I$
over a whole interval, we restrict its Bernstein expansion to that
interval by subdivision; since the Bernstein basis functions are
nonnegative and sum to one, the largest coefficient is an upper bound.
A second evaluation uses the polar form of the same polynomial, with
copies of the two endpoints as arguments. Upper dyadic enclosures of
the positive scores are verified by raising them to their 64th powers,
and the logarithms are enclosed by rational series. The smallest
certified rate exceeds $0.18355862$.

At the isotropic state of the finite-block example of
Ref.~\cite{fang}, $(d,F)=(3,9/10)$, we use the simpler maximum-row
bound.
\providecommand{\FangLower}{1.00344789}
\providecommand{\FangUpper}{1.015863168}
The $20$-copy witness has 12 outcomes, with rational effects and
positive rational scores. Substituting it in Eq.~\eqref{eq:cert-signed}
puts its largest weighted absolute row sum at $j=18$. The resulting
rate lies in $(1.00344789390032,1.00344789390034)$. We checked this enclosure in
two ways, by binomial expansion and by successive multiplication of the
linear factors, using rational upper and lower bounds for each logarithm
in both calculations.

\subsection{An upper bound from harmonic measure}
We now give the finite calculation that sharpens the analytic strict
inequality. In this subsection logarithms without a subscript are
natural. Write $U=\log_2d-h_2(F)-e\log_2(d-1)$, $s=e/[F(d-1)]$ and
$L=\log(1/s)$, and set
\[
 Z(z)=[\Phi_d+(d-1)z(I-\Phi_d)/(d^2-1)]/d.
\]
Write $h(z)=\|Z(z)^\Gamma\|_1$. For any distillation effect $T$ with
output dimension $K$, the periodic subharmonic function
\[
 u(x,y)=n^{-1}\log|2^{nU}\operatorname{Tr}[T Z(\exp(x+iy))^{\otimes n}]|-e x
\]
satisfies $u(-L,y)\le0$, and $u\le\beta_+:=\max(0,U\ln 2-n^{-1}\ln K)$
on $[-L,0]$; both bounds follow from the positivity of the coefficients
and log-convexity between the physical and PPT comparison radii. For
$x\ge0$ there are two upper bounds,
\[
 u(x,y)-\beta_+\le\min\{g(x),\log h(\exp(x+iy))-ex\}.
\]
Here $g(x)=\log[(1+(d-1)\exp x)/d]-ex$. The first bound follows from
the positivity of the coefficients and the fact that the pencil at
positive radius $\exp x\ge1$, divided by its trace, is PPT; the second
is trace-norm duality. Put $g=0$ on $[-L,0]$. As in the analytic
converse, any sequence of tests with acceptance tending to one satisfies
$\liminf_n u(-L+tL,0)\ge0$ at each fixed $0<t<1$. Indeed, some central
error count has acceptance times binomial probability $\exp[-o(n)]$,
and evaluating its monomial at the new positive radius proves the
assertion whenever the fidelity stays bounded away from zero.

Divide $[-L,L]$ into $2M$ equal layers, $x_j=j\ell$ with $\ell=L/M$,
and the angle into $N$ equal cells. We represent nonnegative deficits
$D_j$ by lower bounds on whole cells, so that
$u(x_j,y)\le\beta_++g(x_j)-D_j(y)$.
Start with zero on $j\le0$, and on $j>0$ take the positive part of a
lower enclosure of $\log[(1+(d-1)\exp x_j)/(d h(\exp(x_j+iy)))]$.
Separation of variables, or the strip Poisson formula at the midline,
gives the valid update
\begin{align}
 D_j(y)\leftarrow\max\bigg\{D_j(y),\;& g(x_j)
       -\tfrac12[g(x_{j-1})+g(x_{j+1})]\notag\\[-2pt]
   &+ [K_\ell*(D_{j-1}+D_{j+1})](y)\bigg\},\label{eq:cert-layers}\\
 K_\ell(y)&=\sum_{k\in\mathbb Z}
   {1\over4\ell\cosh[\pi(y+2\pi k)/(2\ell)]}.
\end{align}
The kernel has mass $1/2$, and cellwise lower enclosures of the
convolution suffice: for cells whose centres are $b$ angular steps
apart by the shorter path, keep one image and evaluate the positive
decreasing kernel at distance $(b+1)2\pi/N$. Starting from these
initial values and applying the update any finite number of times
preserves rigorous bounds, so convergence need not be assumed.

In the strip $[-L,0]$, comparison at a fraction $t$ of the width now
gives $0\le t\beta_+-\int k_t(y)D_0(y)\,dy$ in the limit along
fidelities that do not vanish, with the periodized kernel
\begin{equation}
 k_t(y)=\sum_{j\in\mathbb Z}
   {\sin(\pi t)\over 2L[\cosh(\pi(y+2\pi j)/L)+\cos(\pi t)]}.
\end{equation}
Strictly, the inequality concerns the lower limit of $\beta_+$. Letting
$t\downarrow0$ after the coding limit gives a gap in bits of at least
$(\ln2)^{-1}\int\kappa_L(y)D_0(y)\,dy$, where
\begin{equation}
 \kappa_L(y)=\sum_{j\in\mathbb Z}{\pi\over
 2L[\cosh(\pi(y+2\pi j)/L)+1]} .\label{eq:cert-exit}
\end{equation}
This kernel has mass one, and dominated convergence follows from the
exponentially decreasing tails of the summands. In the final
integration it suffices to keep the two nearest images, with lower
cellwise bounds.

The same certificate also shows exponential rejection above its bound.
Write
\[
 C_t=t^{-1}\int k_t(y)D_0(y)\,dy,\qquad C_0=\int\kappa_L(y)D_0(y)\,dy,
\]
so that $C_t\to C_0$ as $t\downarrow0$. Since the deficit bounds hold at
every block length, the strip comparison gives
$u(-L+tL,0)\le t(\beta_+-C_t)$ before any coding limit is taken. If the
rate $\liminf_nn^{-1}\log_2K_n$ exceeds $U-C_0/\ln2$, fix $t>0$ so small
that the right-hand side is at most $-c<0$ for all large $n$. The
argument of Appendix~\ref{app:rejection}, applied
at the radius $r=s\exp(tL)$ in place of $\sqrt s$, then shows that the
fidelity decays exponentially in $n$. The certificate thus supplies an
exponential strong-converse bound in its own right.

For $(3,9/10)$ the update is applied in $1200$ sweeps on the grid
$(N,M)=(4096,24)$, with cell deficits and stencil entries rounded down
in units of $10^{-9}$, integer convolution, and explicit bounds on
intermediate products.

For $(2,3/4)$ the sweeps can be dispensed with altogether. Let $b$ be
the fixed array of initial deficit bounds, and hold the two boundary
layers at their initial values. On the interior layers define the
real-valued map
\[
 \mathcal H(D)_j=\max\bigl\{b_j,\;p_j+\underline K_\ell*(D_{j-1}+D_{j+1})\bigr\},
\]
where $p_j$ and $\underline K_\ell$ are rounded-down enclosures of the
penalty $g(x_j)-\tfrac12[g(x_{j-1})+g(x_{j+1})]$ and of the
convolution stencil in~\eqref{eq:cert-layers}. The first argument of
the maximum is the fixed initial obstacle $b_j$, not the current
iterate. This matters: the update~\eqref{eq:cert-layers}, which retains
$D_j$, is not a contraction, and every array satisfies it trivially.
Because the rounded stencil has total mass strictly less than one half,
$\mathcal H$ is monotone and a strict contraction in the maximum norm.
Its iterates from $b$ are valid deficit bounds and increase to its
unique fixed point, which is therefore valid too. Now let $a\ge0$ have
boundary values at most $b$ and satisfy $a\le\mathcal H(a)$. Raising
its boundary values to $b$ preserves the interior inequality, and
iterating $\mathcal H$ from the result increases to the same fixed
point; hence $a$ lies below that point, and is itself a valid deficit
bound. The map is evaluated in real arithmetic, and rounding its
affine term downwards serves only to verify a sufficient inequality for
a proposed integer array, which it suffices to check cell by cell. We do
so on the grid $(N,M)=(65536,7)$, in units of $10^{-12}$, where the
rounded kernel has mass $0.499694539492$. The array was proposed by
floating-point iteration, but it enters the proof only through the
exact verification. In both cases, independent implementations use
interval elementary functions or fixed-point arithmetic with rational
remainder bounds, and evaluate the convolutions exactly.
The result is
\begin{align}
 0.18355862&\le D_{\rm PPT}(\rho_{2,3/4})<0.188642407,\notag\\
 &\hspace{4mm}U_2=0.18872187554\ldots;\label{eq:cert-qubit}\\
 \FangLower&\le D_{\rm PPT}(\rho_{3,9/10})<\FangUpper,\notag\\
 &\hspace{4mm}U_3=1.01596690713\ldots .\label{eq:cert-qutrit}
\end{align}
These are enclosures of the asymptotic rates, not error bars on
estimates of them. By contrast, Ref.~\cite{fang} worked at fixed error
$10^{-3}$ and fitted block lengths up to 100; such a fit is neither an
asymptotic converse nor a proof of achievability.

\section{Construction of the depolarizing-channel figure}\label{app:figure}
In Fig.~\ref{fig:depol} the horizontal coordinate is $F=1-3p/4$, where $p$ is the replacement parameter of $\mathcal N_p(X)=(1-p)X+p\operatorname{Tr}(X)I/2$. The dashed curve in panel (a) is the analytic strict upper bound~\eqref{sm:strict} with $C=1/8$, and in the notation of Eq.~\eqref{sm:gamma} we take
\begin{align}
 b&=\min\{(1-s)/2,\ e/[2\pi(C+e/8)]\},\nonumber\\
 H&=eb/\pi-(C+e/8)b^2,\qquad a=\sqrt{H/(3C)} .
\end{align}
Throughout $1/2<F<1$ these choices give $0<b<1-s$, $0<a<\pi$ and $\gamma=2H/3>0$. The isolated upper triangle at $F=3/4$ marks the stronger harmonic-measure certificate, not a point of this dashed curve.

The lower curve is the maximum of zero, qubit hashing, and the rates of fixed block witnesses of lengths $9,24,32,94$. For an ordinary witness define
\[
 b_i=m^{-1}\left[\sum_y a_y(i)\log_2 v(y)-\log_2\max_j L(j)\right].
\]
The rate is then $\sum_i\binom mi F^i(1-F)^{m-i}b_i$, as one sees by substituting the physical measurement probabilities into the finite-block bound above. Each of the smaller witnesses uses a single set of scores. For $m=94$ we take the minimum of the ten polynomials obtained from the interval prices in Eq.~\eqref{eq:cert-type}. Since the Bernstein weights are nonnegative, rounding their coefficients downwards preserves the bound at \emph{all} fidelities, so the curve is a genuine lower bound rather than an interpolation between optimized points. The endpoints are fixed by separability and by perfect transmission. Panel (b) subtracts the two fixed-state bounds from $1-h_2(3/4)$; the resulting interval is a rigorous enclosure, not a statistical uncertainty.

The LOCC markers are plotted only at sampled fidelities. The yields of the lookahead protocol and of the protocol of Ref.~\cite{hostens} are taken from Fig.~3 of our Ref.~\cite{lookahead}; those of the protocol of Ref.~\cite{abdelhadi} are the values computed for the same comparison. All include the cost of the local recurrence steps~\cite{bennett,dejmps}. Table~\ref{sm:loccdata} lists the interior values, which are also provided in the ancillary files together with the historical benchmarks shown. They are reported numerical yields of particular protocols, and should not be confused with the interval-certified PPT results.

\par\vspace{0.7\baselineskip}\noindent\begin{minipage}{\linewidth}\makeatletter\def\@captype{table}\makeatother
\caption{Interior LOCC markers in Fig.~\ref{fig:depol}, in ebits per original use: yields of the lookahead protocol of Ref.~\cite{lookahead} and of the benchmarks compared in its Fig.~3. They are numerical protocol yields, not PPT certificates.}\label{sm:loccdata}
\begin{ruledtabular}
\begin{tabular}{cccc}
$F$ & Lookahead~\cite{lookahead} & Ref.~\cite{hostens}, $q=6$ & Ref.~\cite{abdelhadi}\\
0.55 & 0.0002687 & 0.0002374 & 0.0002049\\
0.60 & 0.0033509 & 0.0027253 & 0.0029006\\
0.65 & 0.0127559 & 0.0118018 & 0.0112545\\
0.70 & 0.0341941 & 0.0316614 & 0.0264228\\
0.75 & 0.0880237 & 0.0615569 & 0.0754745\\
0.80 & 0.1662107 & 0.1283103 & 0.1492284\\
0.85 & 0.2762395 & 0.2711329 & 0.2353786\\
0.90 & 0.4431724 & 0.4471561 & 0.4105348\\
0.95 & 0.6656808 & 0.6694701 & 0.6524905\\
\end{tabular}
\end{ruledtabular}
\end{minipage}\par
\end{document}